\documentclass[]{llncs}
\usepackage{algorithm}
\usepackage{algorithmic}
\usepackage{mathrsfs}
\usepackage{multirow}
\usepackage{graphicx,wrapfig,lipsum}
\usepackage{url}
\usepackage{amsfonts}
\usepackage{times}
\usepackage{pgfplots}
\usepackage{tikz}
\usepgfplotslibrary{groupplots}
\usetikzlibrary{shapes,arrows,automata,positioning}
\usepackage[latin1]{inputenc}
\usepackage{verbatim}
\usepackage{makecell}
\usepackage{booktabs}
\usepackage{adjustbox}
\usepackage[compatibility=false]{caption}
\usepackage{subcaption}

\newcommand{\specialcell}[2][c]{%
  \begin{tabular}[#1]{@{}c@{}}#2\end{tabular}}

\begin{document}
\title{Should We Learn Probabilistic Models for Model Checking? A New Approach and an Empirical Study\thanks{This work was supported by NRF Award No.\ NRF2014NCR-NCR001-40.}}

%
%
%
\author{
Jingyi Wang\inst{1}, Jun Sun\inst{1}, Qixia Yuan\inst{2} and Jun Pang\inst{2} \\
\institute{
Singapore University of Technology and Design \\
\email{\{jingyi\_wang@mymail.,sunjun@\}sutd.edu.sg}
\and
University of Luxembourg \\
\email{\{qixia.yuan,jun.pang\}@uni.lu}
}
}


%
\maketitle

\begin{abstract}
Many automated system analysis techniques (e.g., model checking, model-based testing) rely on first obtaining a model of the system under analysis. System modeling is often done manually, which is often considered as a hindrance to adopt model-based system analysis and development techniques. To overcome this problem, researchers have proposed to automatically ``learn'' models based on sample system executions and shown that the learned models can be useful sometimes. There are however many questions to be answered. For instance, how much shall we generalize from the observed samples and how fast would learning converge? Or, would the analysis result based on the learned model be more accurate than the estimation we could have obtained by sampling many system executions within the same amount of time? In this work, we investigate existing algorithms for learning probabilistic models for model checking, propose an evolution-based approach for better controlling the degree of generalization and conduct an empirical study in order to answer the questions. One of our findings is that the effectiveness of learning may sometimes be limited.
\end{abstract}

\section{Introduction}
Many system analysis techniques rely on first obtaining a system model. The model should be accurate and often is required to be at a proper level of abstraction. For instance, model checking~\cite{clarke1999model,baier2008principles} works effectively if the user-provided model captures all the relevant behavior of the system and abstracts away the irrelevant details. With such a model as well as a given property, a model checker would automatically verify the property or falsify it with a counterexample. Alternatively, in the setting of probabilistic model checking (PMC, see  Sect.~\ref{background})~\cite{baier2008principles,bianco1995model}, the model checker would calculate the probability of satisfying the property.

Model checking is perhaps not as popular as it ought to be due to the fact that a good model is required beforehand. For instance, a model which is too general would introduce spurious counterexamples, whereas the checking result based on a model which under-approximates the relevant system behavior is untrustworthy. In the setting of PMC, users are required to provide a probabilistic model (e.g., a Markov chain~\cite{baier2008principles}) with accurate probabilistic distributions, which is often challenging.

In practice, system modeling is often done manually, which is both time-consuming and error-prone. Worse, it could be infeasible if the system is a black box or it is so complicated that no accurate model is known (e.g., the chemical reaction in a water treatment system~\cite{swat}). This is often considered by industry as one hindrance to adopt otherwise powerful techniques like model checking. Alternative approaches which would rely less on manual modeling have been explored in different settings. One example is statistical model checking (SMC, see Sect.~\ref{background})~\cite{younes2002probabilistic,sen2004statistical}. The main idea is to provide a statistical measure on the likelihood of satisfying a property, by observing sample system executions and applying standard techniques like hypothesis testing~\cite{bauer2006monitoring,havelund2002synthesizing,younes2002probabilistic}. SMC is considered useful partly because it can be applied to black-box or complex systems when system models are not available. 

Another approach for avoiding manual modeling is to automatically learn models. A variety of learning algorithms have been proposed to learn a variety of models, e.g.,~\cite{sen2004learning,ron1996power,carrasco1994learning,de2010grammatical}. It has been showed that the learned models can be useful for subsequent system analysis in certain settings, especially so when having a model is a must. Recently, the idea of model learning has been extended to system analysis through model checking. In~\cite{mao2011learning,chen2012learning,mao2012learning}, it is proposed to learn a probabilistic model first and then apply techniques like PMC to calculate the probability of satisfying a property based on the learned model. On one hand, learning is beneficial, and it solves some known drawbacks of SMC or even simulation-based system analysis methods in general. 
For instance, since SMC relies on sampling \emph{finite} system executions, it is challenging to verify un-bounded properties~\cite{younes2011statistical,rohr2013simulative}, whereas we can verify un-bounded properties based on the learned model through PMC. Furthermore, the learned model can be used to facilitate other system analysis tasks like model-based testing and software simulation for complicated systems. On the other hand, learning essentially is a way of generalizing the sample executions and there are often many variables. It is thus worth investigating how the sample executions are generalized and whether indeed such learning-based approaches are justified.

In particular, we would like to investigate the following research questions. Firstly, how can we control the degree of generalization for the best learning outcome, since it is known that both over-fitting or under-fitting would cause problems in subsequent analysis? Secondly, often it is promised that the learned model would converge to an accurate model of the original system, if the number of sample executions is sufficiently large. In practice, there could be only a limited number of sample executions and thus it is valid to question how fast the learning algorithms converge. Furthermore, do learning-based approaches offer better analysis results if alternative approaches which do not require a learned model, like SMC, are available?

In order to answer the above questions, we mainly make the following contributions. Firstly, we propose a new approach (Sect.~\ref{learning}) to better control the degree of generalization than existing approaches (Sect.~\ref{existing}) in model learning. The approach is inspired by our observations on the limitations of existing learning approaches. Experiment results show that our approach converges faster 
than 
existing approaches while providing better or similar analysis results.
	 Secondly, we develop a software toolkit \textsc{Ziqian}, realizing previously proposed learning approaches for PMC as well as our approach so as to systematically study and compare them in a fair way. Lastly, we conduct an empirical study on comparing different model learning approaches against a suite of benchmark systems, two real world systems, as well as randomly generated models (Sect.~\ref{evaluation}). One of our findings suggests that learning models for model checking might not be as effective compared to SMC given the same time limit. However, the learned models may be useful when manual modeling is impossible.
From a broader point of view, our work is a first step towards investigating the recent trend on adopting machine learning techniques to solve software engineering problems. We remark there are extensive existing research on learning non-probabilistic models (e.g.,~\cite{angluin1987learning}), which is often designed for different usage and is thus beyond the scope of this work. We review related work and conclude this paper in Sect.~\ref{relatedwork}.


\section{Preliminary} \label{background}
In this work, the model that we focus on is discrete-time Markov chains (DTMC)~\cite{baier2008principles}. The reason is that most existing learning algorithms generate DTMC and it is still ongoing research on how to learn other probabilistic models like Markov Decision Processes~\cite{brazdil2014verification,mao2012learning,sen2004learning,mao2011learning,chen2012learning}. Furthermore, the learned DTMC is aimed for probabilistic analysis by methods like PMC, among others. In the following, we briefly introduce DTMC, PMC as well as SMC so that we can better understand the context.\\

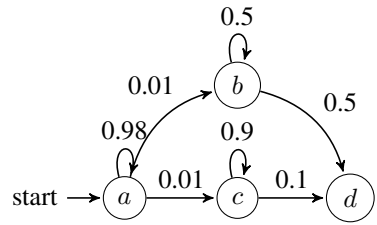
\begin{wrapfigure}{r}{5cm}
	 \vspace{-8mm}
 \begin{tikzpicture}[->,>=stealth',shorten >=1pt,auto,node distance=1.5cm,
semithick]
    \tikzset{state/.style={circle,draw,thin,minimum size=0.4cm}},
\tikzstyle{every state}=[text=black]

\node[initial,state] (s1) {$a$};
\node[state] (s3) [right of=s1] {$c$};
\node[state] (s2) [above of=s3] {$b$};
\node[state] (s4) [right of=s3] {$d$};

\path (s1) edge [loop above] node {0.98} (s1)
edge [bend left, pos=0.75] node {0.01} (s2)
edge node {0.01} (s3)
(s2) edge [loop above] node {0.5} (s2)
edge [bend left] node {0.5} (s4)
(s3) edge [loop above] node {0.9} (s3)
edge node {0.1} (s4);
\end{tikzpicture}
 \caption{DTMC of \emph{egl} protocol.}
  \vspace{-8mm}
  \label{fig:dtmc}
 \end{wrapfigure}

\noindent \textbf{Markov Chain} A DTMC $\mathcal{D}$ is a triple tuple $(S,\imath_{init},Tr)$, where $S$ is a countable, nonempty set of states; $\imath_{init}: S \to [0,1]$ is the initial distribution s.t.~$\sum_{s\in S}\imath_{init}(s)=1$; 
 and $Tr: S \times S \to [0,1]$ is the transition probability assigned to every pair of states which satisfies the following condition: $\sum_{s'\in S}Tr(s,s')=1$. $\mathcal{D}$ is \emph{finite} if $S$ is finite. For instance, an example DTMC modelling the \emph{egl} protocol~\cite{KNP12b} is shown in Figure~\ref{fig:dtmc}.

A DTMC induces an underlying digraph where states are vertices and there is an edge from $s$ to $s'$ if and only if $Tr(s,s')>0$. \emph{Paths} of DTMCs are maximal paths in the underlying digraph, defined as infinite state sequences $\pi=s_0s_1s_2\cdots\in S^{\omega}$ such that $Tr(s_i,s_{i+1})>0$ for all $i\geq 0$. We write $\mathit{Path}^{\mathcal{D}}(s)$ to denote the set of all infinite paths of $\mathcal{D}$ starting from state $s$. \\

\noindent \textbf{Probabilistic Model Checking} PMC~\cite{bianco1995model,baier2008principles} is a formal analysis technique for stochastic systems including DTMCs. Given a DTMC $\mathcal{D} = (S,\imath_{init},Tr)$ and a set of propositions $\Sigma$, we can define a function $L: S \rightarrow \Sigma$ which assigns valuation of the propositions in $\Sigma$ to each state in $S$.
Once each state is labeled, given a path in $\mathit{Path}^{\mathcal{D}}(s)$, we can obtain a corresponding sequence of propositions labeling the states.

Let $\Sigma^\star$ and $\Sigma^\omega$ be the set of all finite and infinite strings over $\Sigma$ respectively. A property of the DTMC can be specified in temporal logic. Without loss of generality, we focus on Linear Time Temporal logic (LTL) and probabilistic LTL in this work. An LTL formula $\varphi$ over $\Sigma$ is defined by the syntax: \\
\centerline{$\varphi::= true\ | \ \sigma\ |\ \varphi_1\land \varphi_2\ |\ \neg\varphi\ |\ \textbf{X}\varphi\ |\ \varphi_1 \textbf{U} \varphi_2$}\\
where $\sigma \in \Sigma$ is a proposition; $\textbf{X}$ is intuitively read as `next' and $\textbf{U}$ is read as `until'.
We remark commonly used temporal operators like $\textbf{F}$ (which reads `eventually') and $\textbf{G}$ (which reads `always') can be defined using the above syntax, e.g., $\textbf{F} \varphi$ is defined as $true \textbf{U} \varphi$. Given a string $\pi$ in $\Sigma^\star$ or $\Sigma^\omega$, we define whether $\pi$ satisfies a given LTL formula $\varphi$ in the standard way~\cite{baier2008principles}.

Given a path $\pi$ of a DTMC, we write $\pi \models \varphi$ to denote that the sequence of propositions obtained from $\pi$ satisfies $\varphi$ and $\pi \not \models \varphi$ otherwise.
Furthermore, a probabilistic LTL formula $\phi$ of the form $Pr_{\bowtie r}(\varphi)$ can be used to quantify the probability of a system satisfying the LTL formula $\varphi$, where $\bowtie \in \{ \geq,\leq,=\}$ and $r \in [0,1]$ is a probability threshold. A DTMC $\mathcal{D}$ satisfies $Pr_{\bowtie r}(\varphi)$ if and only if the accumulated probability of all paths obtained from the initial state of $\mathcal{D}$ which satisfy $\varphi$ satisfies the condition $\bowtie r$. Given a DTMC $\mathcal{D}$ and a probabilistic LTL property $Pr_{\bowtie r}(\varphi)$, the PMC problem can be solved using methods like the automata-theoretic approach~\cite{baier2008principles}. 
We skip the details of the approach and instead remark that the complexity of PMC is doubly exponential in the size of $\varphi$ and polynomial in the size of $\mathcal{D}$. \\

\noindent \textbf{Statistical Model Checking}
SMC is a Monte Carlo method to solve the probabilistic verification problem based on system simulations. Its biggest advantage is perhaps that it does not require the availability of system models~\cite{clarke2011statistical}. SMC works by sampling system behaviors randomly (according to certain underlying probabilistic distribution) and observing how often a given property $\varphi$ is satisfied. The idea is to provide a statistical measure on the likelihood of satisfying $\varphi$ based on the observations, by applying techniques like hypothesis testing~\cite{bauer2006monitoring,havelund2002synthesizing,younes2002probabilistic}.
We
refer readers to~\cite{younes2002probabilistic,baier2008principles} for details.

\section{Probabilistic Model Learning} \label{existing}
Learning models from sample system executions for the purpose of PMC has been explored extensively in recent years~\cite{sen2004learning,ron1996power,carrasco1994learning,de2010grammatical,mao2011learning,chen2012learning,mao2012learning}. In this section, we briefly present 
 existing model learning algorithms for two different settings. 
\vspace{-2mm}
\subsection{Learn from Multiple Executions}\label{learning1}
In the setting that the system can be reset and restarted multiple times, a set of independent executions of the system can be collected as input for learning. Learning algorithms in this category make the following assumptions~\cite{mao2011learning}. First, the underlying system can be modeled as a DTMC. Second, the sampled system executions are mutually independent. Third, the length of each simulation is independent. 

Let $\Sigma$ denote the alphabet of the system observations such that each letter $e \in \Sigma$ is an observation of the system state. A system execution is then a finite string over $\Sigma$.
The input in this setting is a finite set of strings $\Pi \subseteq \Sigma^{\star}$. For any string $\pi \in \Sigma^{\star}$, let $\mathit{prefix}(\pi)$ be the set of all prefixes of $\pi$ including the empty string $\langle \rangle$. Let $\mathit{prefix}(\Pi)$ be the set of all prefixes of any string $\pi \in \Pi$. The set of strings $\Pi$ can be naturally organized into a tree $\mathit{tree}(\Pi) = (N, root, E)$ where each node in $N$ is a member of $\mathit{prefix}(\Pi)$; the root is the empty string $\langle \rangle$; and $E \subseteq N \times N$ is a set of edges such that $(\pi, \pi')$ is in $E$ if and only if there exists $e \in \Sigma$ such that $\pi \cdot \langle e \rangle = \pi'$ where $\cdot$ is the sequence concatenation operator.


The idea of the learning algorithms is to generalize $tree(\Pi)$ by merging the nodes according to certain criteria in certain fixed order. Intuitively, two nodes should be merged if they are likely to represent the same state in the underlying DTMC. Since we do not know the underlying DTMC, whether two states should be merged is decided through a procedure called \emph{compatibility test}. We remark the compatibility test effectively controls the degree of generalization. Different types of compatibility test have been studied~\cite{carrasco1994learning,ron1995learnability,kermorvant2002stochastic}. 
We present in detail the compatibility test adopted in the AALERGIA algorithm~\cite{mao2011learning} as a representative. First, each node $\pi$ in $\mathit{tree}(\Pi)$ is labeled with the number of strings $\mathit{str}$ in $\Pi$ such that $\pi$ is a prefix of $\mathit{str}$. Let $L(\pi)$ denote its label. Two nodes $\pi_1$ and $\pi_2$ in $\mathit{tree}(\Pi)$ are considered compatible if and only if they satisfy two conditions. The first condition is $\mathit{last}(\pi_1) = \mathit{last}(\pi_2)$ where $\mathit{last}(\pi)$ is the last letter in a string $\pi$, i.e., if the two nodes are to be merged, they must agree on the last observation (of the system state). The second condition is that the future behaviors from $\pi_1$ and $\pi_2$ must be sufficiently similar (i.e., within Angluin's bound~\cite{angluin1988identifying}). Formally, given a node $\pi$ in $\mathit{tree}(\Pi)$, we can obtain a probabilistic distribution of the next observation by \emph{normalizing} the labels of the node and its children. In particular, for any event $e \in \Sigma$, the probability of going from node $\pi$ to $\pi \cdot \langle e \rangle$ is defined as: $Pr(\pi, \langle e \rangle) = \frac{L(\pi \cdot \langle e \rangle)}{L(\pi)}$. We remark the probability of going from node $\pi$ to itself is $Pr(\pi,\langle\rangle)=1-\sum_{e\in\Sigma}Pr(\pi,\langle e\rangle)$, i.e., the probability of not making any more observation. The multi-step probability from node $\pi$ to $\pi \cdot \pi'$ where $\pi' = \langle e_1, e_2, \cdots, e_k \rangle$, written as $\mathit{Pr}(\pi, \pi')$, is the product of the one-step probabilities:
\begin{equation}
		\mathit{Pr}(\pi, \pi') = \mathit{Pr}(\pi, \langle e_1 \rangle) \times \mathit{Pr}(\pi \cdot \langle e_1 \rangle, \langle e_2 \rangle)\times \\
		 \cdots \times \mathit{Pr}(\pi \cdot \langle e_1, e_2, \cdots, e_{k-1} \rangle, \langle e_k \rangle)
	\end{equation}
Two nodes $\pi_1$ and $\pi_2$ are compatible if the following is satisfied:
\begin{equation}
		|\mathit{Pr}(\pi_1, \pi) - \mathit{Pr}(\pi_2, \pi)|< \sqrt{6 \epsilon \log(L(\pi_1))/L(\pi_1)} + \\
		\sqrt{6 \epsilon \log(L(\pi_2))/L(\pi_2)}
	\end{equation}
for all $\pi \in \Sigma^{\star}$. We highlight that $\epsilon$ used in the above condition is a parameter which effectively controls the degree of state merging. Intuitively, a larger $\epsilon$ leads to more state merging, thus fewer states in the learned model.

If $\pi_1$ and $\pi_2$ are compatible, the two nodes are merged, i.e., the tree is transformed such  that the incoming edge of $\pi_2$ is directed to $\pi_1$. Next, for any $\pi \in \Sigma^*$, $L(\pi_1 \cdot \pi)$ is incremented by $L(\pi_2 \cdot \pi)$. The algorithm works by iteratively identifying nodes which are compatible and merging them until there are no more compatible nodes. 
After merging all compatible nodes, the last phase of the learning algorithms in this category is to normalize the tree so that it becomes a DTMC. 

\vspace{-2mm}
\subsection{Learn from a Single Execution}\label{learning2}
In the setting that the system cannot be easily restarted, e.g., real-world cyber-physical systems. We are limited to observe the system for a long time and collect a single, long execution as input. Thus, the goal is to learn a model describing the long-run, stationary behavior of a system, in which system behaviors are decided by their finite variable length memory of the past behaviors.

In the following, we fix $\alpha$ to be the single system execution. Given a string $\pi = \langle e_0, e_1, \cdots, e_k \rangle$, we write $\mathit{suffix}(\pi)$ to be the set of all suffixes of $\pi$, i.e., $\mathit{suffix}(\pi)=\{\langle e_i, \cdots, e_k \rangle |0 \leq i \leq k\}\cup\{\langle \rangle\}$. Learning algorithms in this category~\cite{chen2012learning,ron1996power} similarly construct a tree $\mathit{tree}(\alpha) = (N, root, E)$ where $N$ is the set of suffixes of $\alpha$; $\mathit{root} = \langle \rangle$; and there is an edge $(\pi_1, \pi_2) \in E$ if and only if $\pi_2 = \langle e \rangle \cdot \pi_1$. For any string $\pi$, let $\#(\pi, \alpha)$ be the number of times $\pi$ appears as a substring in $\alpha$. A node $\pi$ in $\mathit{tree}(\alpha)$ is associated with a function $Pr_{\pi}$ such that $Pr_{\pi}(e)= \frac{\#(\pi\cdot\langle e \rangle, \alpha)}{\#(\pi, \alpha)}$ for every $e\in\Sigma$, which is the likelihood of observing $e$ next given the previous observations $\pi$. Effectively, function $Pr_{\pi}$ defines a probabilistic distribution of the next observation.


Based on different suffixes of the execution, different probabilistic distributions of the next observation will be formed. For instance, the probabilistic distribution from the node $\langle e \rangle$ where $e$ is the last observation would predict the distribution only based on the last observation, whereas the node corresponding to the sequence of all previous observations would have a prediction based the entire history. The central question is how far we should look into the past in order to predict the future. As we observe more history, we will make a better prediction of the next observation. Nonetheless, constructing the tree completely (no generalization) is infeasible 
and the goal of the learning algorithms is thus to grow a part of the tree which would give a ``good enough'' prediction by looking at a small amount of history. The questions are then: what is considered ``good enough'' and how much history is necessary. The answers control the degree of generalization in the learned model.

In the following, we present the approach in~\cite{chen2012learning} as a representative of algorithms proposed in the setting. Let $\mathit{fre}(\pi, \alpha) = \frac{\#(\pi, \alpha)}{|\alpha|-|\pi|-1}$ where $|\pi|$ is the length of $\pi$ be the relative frequency of having substring $\pi$ in $\alpha$. Algorithm~\ref{alg:pstlearning} shows the algorithm for identifying the right tree by growing it on-the-fly. Initially, at line 1, the tree $T$ contains only the root $\langle \rangle$. Given a threshold $\epsilon$, we identify the set $S = \{\pi|\mathit{fre}(\pi, \alpha) > \epsilon\}$ at line 2, which are substrings appearing often enough in $\alpha$ and are candidate nodes to grow in the tree. The loop from line 3 to 7 keeps growing $T$. In particular, given a candidate node $\pi$, we find the longest suffix $\pi'$ in $T$ at line 4 and if we find that adding $\pi$ would improve the prediction of the next observations by at least $\epsilon$, $\pi$ is added, along with all of its suffixes if they are currently missing from the tree (so that we maintain all suffixes of all nodes in the tree all the time). Whether we add node $\pi$ into tree $T$ or not, we update the set of candidate $S$ to include longer substrings of $\alpha$ at line 6. When Algorithm~\ref{alg:pstlearning} terminates, the tree contains all nodes which would make a \emph{good enough} prediction. Afterwards, the tree is transformed into a DTMC where
the leafs of $\mathit{tree}(\alpha)$ are turned into states in the DTMC (refer to~\cite{ron1996power} for details). 


\begin{algorithm}[t]
\caption{$Learn\ PST$}\label{alg:pstlearning}
\begin{algorithmic}[1]
\STATE Initialize $T$ to be a single root node representing $\langle\rangle$; \\
\STATE Let $S = \{\sigma|fre(\sigma, \alpha) > \epsilon\}$ be the candidate suffix set; \\
\WHILE{$S$ is not empty}
\STATE Take any $\pi$ from $S$; Let $\pi'$ be the longest suffix of $\pi$ in $T$; \\
\STATE (B) If $\mathit{fre}(\pi,\alpha) \cdot \sum_{\sigma\in\Sigma} \mathit{Pr}(\pi, \sigma) \cdot \log{\frac{\mathit{Pr}(\pi, \sigma)}{\mathit{Pr}(\pi', \sigma)}}\ge \epsilon$ \\
~~~~add $\pi$ and all its suffixes which are not in $T$ to $T$;
\STATE (C) If $\mathit{fre}(\pi,\alpha) > \epsilon$, add $\langle e \rangle \cdot \pi$ to $\mathcal{S}$ for every $e\in\Sigma$ if $\mathit{fre}(\langle e \rangle \cdot \pi, \alpha)>0$;
\ENDWHILE
\end{algorithmic}
\end{algorithm}


\section{Learning through Evolution}~\label{learning}
Model learning essentially works by generalizing the sample executions. The central question is thus how to control the degree of generalization. 
To find the best degree of generalization, both~\cite{mao2011learning} and~\cite{chen2012learning} proposed to select the `optimal' $\epsilon$ value using the golden section search of the highest Bayesian Information Criterion (BIC) score. For instance, in~\cite{mao2011learning}, the BIC score of a learned model $M$, given the sample executions $\Pi$, is computed as follows: $log(Pr_{M}(\Pi))- \mu \times |M| \times log(|\Pi|)$ where $|M|$ is the number of states in $M$; $\Pi$ is the total number of observations and $\mu$ is a constant (set to be 0.5 in~\cite{mao2011learning}) which controls the relative importance of the size of the learned model.
This kind of approach to optimize BIC is based on the assumption that the BIC score is a concave function of the parameter $\epsilon$.
Our empirical study (refer to details in section~\ref{evaluation}), however, shows that this assumption is flawed and the BIC score can fluctuate with $\epsilon$. 


In the following, we propose an alternative method for learning models based on genetic algorithms (GA)~\cite{Holland:1992:ANA:129194}. The method is designed to select the best degree of generalization without the assumption of BIC's concaveness. The idea is that instead of using a predefined $\epsilon$ value to control the degree of generalization, we systematically generate candidate models and select the ones using the principle of natural selection so that the ``fittest'' model is selected eventually. In the following, we first briefly introduce the relevant background on GA and then present our approach in detail.
\vspace{-2mm}
\subsection{Genetic Algorithms}
GA~\cite{Holland:1992:ANA:129194} are a set of optimization algorithms inspired by the ``survival of the fittest'' principle of Darwinian theory of natural selection. Given a specific problem whose solution can be encoded as a chromosome, a genetic algorithm typically works in the following steps~\cite{watchmaker}. 
First, an initial population (i.e., candidate solutions) is created either randomly or hand-picked based on certain criteria. Second, each candidate is evaluated using a pre-defined fitness function to see how good it is. Third, those candidates with higher fitness scores are selected as the parents of the next generation. Fourth, a new generation is generated by genetic operators, which either randomly alter (a.k.a.~mutation) or combine fragments of their parent candidates (a.k.a.~cross-over). Lastly, step 2-4 are repeated until a satisfactory solution is found or some other termination condition (e.g., timeout) is satisfied. GA are especially useful in providing approximate `optimal' solutions when other optimization techniques do not apply or are too expensive, or the problem space is too large or complex.


GA are suitable for solving our problem of learning DTMC because we view the problem as finding an optimal DTMC model which not only maximizes the likelihood of the observed system executions but also satisfies additional constrains like having a small number of states. To apply GA to solve our problem, we need to develop a way of encoding candidate models in the form of chromosomes, define operators such as mutation and crossover to generate new candidate models, and define the fitness function to selection better models.
In the following, we present the details of the steps in our approach.
\vspace{-2mm}
\subsection{Learn from Multiple Executions} \label{evolution1}
We first consider the setting where multiple system executions are available. Recall that in this setting, we are given a set of strings $\Pi$, from which we can build a tree representation $tree(\Pi)$. Furthermore, a model is learned through merging the nodes in $tree(\Pi)$. The space of different ways of merging the nodes thus corresponds to the potential models to learn. Our goal is to apply GA to search for the best model in this space. In the following, we first show how to encode different ways of merging the nodes as chromosomes.

Let the size of $tree(\Pi)$ (i.e., the number of nodes) be $X$ and let $Z$ be the number of states in the learned model. A way of merging the nodes is a function which maps each node in $tree(\Pi)$ to a state in the learned model. That is, it can be encoded as a chromosome in the form of a sequence of integers $\langle I_1,I_2,\cdots,I_X \rangle$ where $1 \leq I_i \leq Z$ for all $i$ such that $1 \leq i \leq X$. Intuitively, the number $I_i$ means that node $i$ in $tree(\Pi)$ is mapped into state $I_i$ in the learned model. Besides, the encoding is done such that infeasible models are always avoided. Recall that two nodes $\pi_1$ and $\pi_2$ can be merged only if $last(\pi_1) = last(\pi_2)$, which means that two nodes with different last observation should not be mapped into the same state in the learned model. Thus, we first partition the nodes into $|\Sigma|$ groups so that all nodes sharing the same last observation are mapped to the same group of integers. A chromosome is then generated such that only nodes in the same group can possibly be mapped into the same state. The initial population is generated by randomly generating a set of chromosomes this way. \emph{We remark that in this way all generated chromosomes represent a valid DTMC model.}

\begin{algorithm}[t]
	\caption{Model learning by GA from multiple executions}
	\label{alg:ga1}	
	\begin{algorithmic}[1]
		\REQUIRE $\it{tree}(\Pi)$ and the alphabet $\Sigma$
		\ENSURE A chromosome encoding a DTMC $\mathcal{D}$
		\STATE Let $Z$ be $|\Sigma|$;\ Let $Best$ be $null$;\
		\REPEAT
		\STATE Let $population$ be an initial population with $Z$ states;\
		\STATE Let $generation$ be $1$;\
		\REPEAT
		\STATE Let $newBest$ be the fittest in $population$;\
		\IF{$newBest$ is fitter than $Best$}
		\STATE Set $Best$ to be $newBest$;\
		\ENDIF
		\FORALL{fit pairs $(p_1, p_2)$ in $population$}
		\STATE Crossover $(p_1, p_2)$ to get children $C_1$ and $C_2$;\
		\STATE Mutate $C_1$ and $C_2$;\
		\STATE Add $C_1$ and $C_2$ into $population$;
        \STATE Remove $(p_1, p_2)$ from $population$;
		\ENDFOR
		\STATE $generation\gets generation+1;$
		\UNTIL{$generation>someThreshold$}
		\STATE $Z\gets Z+1;$
		\UNTIL{$Best$ is not improved}
		\RETURN $Best$
	\end{algorithmic}
\end{algorithm}




Formally, the chromosome $\langle I_1,I_2,\cdots,I_X \rangle$ represents a DTMC $M = (S,\imath_{init},Tr)$ where $S$ is a set of $Z$ states. Each state $s$ in $S$ corresponds to a set of nodes in $tree(\Pi)$. Let $nodes(s)$ denote that set. $Tr$ is defined such that for all states $s$ and $s'$ in $M$,
\begin{equation}
 Tr(s,s')=\frac{\sum_{x\in nodes(s)}\sum_{e\in\Sigma | \langle s, e \rangle \in nodes(s')}L(x \cdot \langle e \rangle)}{\sum_{x\in nodes(s)}L(x)}
 \end{equation}
The initial distributions $\imath_{init}$ is defined such that for any state $s \in S$, $\imath_{init}(s) = \sum_{x\in nodes(s)}L(x)/L(\langle\rangle)$.


Next, we define the fitness function.
Intuitively, a chromosome is good if the corresponding DTMC model $M$ maximizes the probability of the observed sample executions and the number of states in $M$ is small. We thus define the fitness function of a chromosome as:
$log(Pr_{M}(\Pi))-\mu \times |M| \times log|\Pi|$ where $|M|$ is the number of states in $M$ and $|\Pi|$ is the total number of letters in the observations and $\mu$ is a constant which represents how much we favor a smaller model size. The fitness function, in particular, the value of $\mu$, controls the degree of generalization. If $\mu$ is 0, $tree(\Pi)$ would be the resultant model; whereas if $\mu$ is infinity, a model with one state would be generated. We remark that this fitness function is the same as the formula for computing the BIC score in~\cite{mao2011learning}. Compared to existing learning algorithms, controlling the degree of generalization in our approach is more intuitive (i.e., different value of $\mu$ has a direct effect on the learned model). In particular, a single parameter $\mu$ is used in our approach, whereas in existing algorithms~\cite{mao2011learning,chen2012learning}, a parameter $\mu$ is used to select the value of $\epsilon$ (based on a false assumption of the BIC being concave), which in turn controls the degree of generalization. From a user point of view, it is hard to see the effect of having a different $\epsilon$ value since it controls whether two nodes are merged in the intermediate steps of the learning process.

Next, we discuss how candidate models with better fitness score are selected. Selection directs evolution towards better models by keeping good chromosomes and weeding out bad ones based on their fitness. Two standard selection strategies are applied. One is \emph{roulette wheel selection.} Suppose $f$ is the average fitness of a population. For each individual $M$ in the population, we select $f_M/f$ copies of $M$. The other is \emph{tournament selection.} Two individuals are chosen randomly from the population and a tournament is staged to determine which one gets selected. The tournament is done by generating a random number $r$ between zero and comparing it to a pre-defined number $p$ (which is larger than $0.5$). If $r$ is smaller than $p$, the individual with a higher fitness score is kept. We refer the readers to~\cite{Holland:1992:ANA:129194} for discussion on the effectiveness of these selection strategies.

After selection, genetic operators like mutation and crossover are applied to the selected candidates. Mutation works by mapping a random node to a new number from the same group, i.e., merging the node with other nodes with the same last observation. For crossover, chromosomes in the current generation are randomly paired and two children are generated to replace them. Following standard approaches~\cite{Holland:1992:ANA:129194}, we adopt three crossover strategies.
\begin{itemize}
	\item{\emph{One-point Crossover.} A crossover point is randomly chosen, one child gets its prefix from the father and suffix from the mother. Reversely for the other child.}
	\item{\emph{Two-point Crossover.} Two crossover points are randomly chosen, which results in two crossover segments in the parent chromosomes. The parents exchange their crossover segments to generate two children.}
	\item{\emph{Uniform Crossover.} One child gets its odd bit from father and even bit from mother. Reversely for the other child.}
\end{itemize}
We remark that during mutation or crossover, we guarantee that only chromosomes representing valid DTMC models are generated, i.e., only two nodes with the same last observations are mapped to the same number (i.e., a state in the learned model).

The details of our GA-based algorithm is shown as Algorithm~\ref{alg:ga1}. Variable $Z$ is the number of states in the learned model. We remark that the number of states in the learned model $M$ is unknown in advance. However, it is at least the number of letters in alphabet $\Sigma$, i.e., when all nodes in $\it{tree}(\Pi)$ sharing the same last observation are merged. Since a smaller model is often preferred, the initial population is generated such that each of the candidate models is of size $|\Sigma|$. The size of the model is incremented by $1$ after each round of evolution. Variable $Best$ records the fittest chromosome generated so far, which is initially set to be $null$ (i.e., the least fit one). At line 3, an initial population of chromosome with $Z$ states are generated as discussed above. The loop from line 5 to 17 then lets the population evolve through a number of generations, during which crossover and mutations take place. At line 18, we then increase the number of states in the model in order to see whether we can generate a fitter chromosome. We stop the loop from line 2 to 19 when the best chromosome is not improved after increasing the number of states. Lastly, the fittest chromosome $Best$ is decoded to a DTMC and presented as the learned model. \\

\noindent \textbf{Example} We use an example to illustrate how the above approach works. For simplicity, assume we have the following collection of executions $\Pi = \{\langle aacd \rangle, \langle abd \rangle, \langle acd \rangle\}$ from the model shown in Figure~\ref{fig:dtmc}. There are in total 10 prefixes of these execution (including the empty string). As a result, the tree $tree(\Pi)$ contains 10 nodes. Since the alphabet $\{a,b,c,d\}$ has size 4, the nodes (except the root) are partitioned into 4 groups so that all nodes in the same group have the same last observation. The initial population contains a single model with 4 states, where all nodes in the same groups are mapped into the same state. After one round of evolution, models with 5 states are generated (by essentially splitting the nodes in one group to two states) and evaluated with the fitness function. The evolution continues until the fittest score does not improve anymore when we add more states.
\vspace{-2mm}
\subsection{Learn from Single Execution}
In the following, we describe our GA-based learning if there is only one system execution.
Recall that we are given a single long system observation $\alpha$ in this setting. The goal is to identify the \emph{shortest dependent history memory} that yields the most precise probability distribution of the system's next observation. That is, we aim to construct a part of $tree(\alpha)$ which transforms to a ``good'' DTMC.
A model thus can be defined as an assignment of each node in $tree(\alpha)$ to either true or false. Intuitively, a node is assigned true if and only if it is selected to predict the next observation, i.e., the corresponding suffix is kept in the tree which later is used to construct the DTMC model. A chromosome (which encodes a model) is thus in the form of a sequence of boolean variable $\langle B_1,B_2,\cdots,B_m \rangle$ where $B_i$ represents whether the $i$-th node is to be kept or not. We remark that not every valuation of the boolean variables is considered a valid chromosome. By definition, if a suffix $\pi$ is selected to predict the next observation, all suffixes of $\pi$ are not selected (since using a longer memory as in $\pi$ predicts better) and therefore their corresponding value must be false. During mutation and crossover, we only generate those chromosomes satisfying this condition so that only valid chromosomes are generated. 

%

A chromosome defined above encodes a part of $tree(\alpha)$, which can be transformed into a DTMC following the approach in~\cite{ron1996power}. Let $M$ be the corresponding DTMC.
The fitness function is defined similarly as in Section~\ref{evolution1}. We define the fitness function of a chromosome as $log(Pr_{M}(\alpha))-\mu \times |M| \times log(|\alpha|)$ where $Pr_{M}(\alpha)$ is the probability of exhibiting $\alpha$ in $M$, $\mu$ is a constant that controls the weight of model size, and $|\alpha|$ is the size of the input execution. Mutation is done by randomly selecting one boolean variable from the chromosome and flip its value. Notice that afterwards, we might have to flip the values of other boolean values so that the chromosome is valid. We skip the discussion on selection and crossover as they are the same as described in Section~\ref{evolution1}.

We remark that, compared to existing algorithms in learning models~\cite{mao2011learning,chen2012learning,mao2012learning}, it is straightforward to argue that the GA-based approaches for model learning do not rely on the assumption needed for BIC. Furthermore, the learned model improves monotonically through generations.


\section{Empirical Study}~\label{evaluation}
The above mentioned learning algorithms are implemented in a self-contained tool called \textsc{Ziqian} (available at~\cite{tool}, approximately 6K lines of Java code). In this work, since the primary goal of learning the models is to verify properties over the systems, we evaluate the learning algorithms by checking whether we can reliably verify properties based on the learned model, by comparing verification results based on the learned models and those based on the actual models (if available). All results are obtained using PRISM~\cite{kwiatkowska2002prism} on a 2.6 GHz Intel Core i7 PC running OSX with 8 GB memory. The constant $\mu$ in the fitness function of learning by GA is set to 0.5. 

Our test objects can be categorized in two groups. The first group contains all systems (\emph{brp}, \emph{lse}, \emph{egl}, \emph{crowds}, \emph{nand}, and \emph{rsp}) from the PRISM benchmark suite for DTMCs~\cite{KNP12b} and a set of randomly generated DTMC models (\emph{rmc}) using an approach similar to the approach in~\cite{DBLP:conf/lpar/TabakovV05}. We refer the readers to~\cite{KNP12b} for details on the PRISM models as well as the properties to be verified. For these models, we collect multiple executions. The second group contains two real-world systems, from which we collect a single long execution. One is the probabilistic boolean networks (\emph{PBN}), which is a modeling framework widely used to model gene regulatory networks (GRNs)~\cite{SDZ02}. In \emph{PBN}, a gene is modeled with a binary valued node and the interactions between genes are expressed by Boolean functions. For the evaluation, we generate random PBNs with 5, 8 and 10 nodes respectively using the tool {\sf ASSA-PBN}~\cite{assa}. 
 The other is a real-world raw water purification system called the Secure Water Testbed (\emph{SWaT})~\cite{swat}. \emph{SWaT} is a complicated system which involves a series of water treatments like ultrafiltration, chemical dosing, dechlorination through an ultraviolet system, etc. 
 We regard \emph{SWaT} as a representative complex system for which learning is the only way to construct a model. Our evaluation consists of the following parts (all models as well as the detailed results are available at~\cite{evaluation}).

\begin{wrapfigure}{r}{6cm}
	 \vspace{-8mm}
 \includegraphics[width=0.5\textwidth]{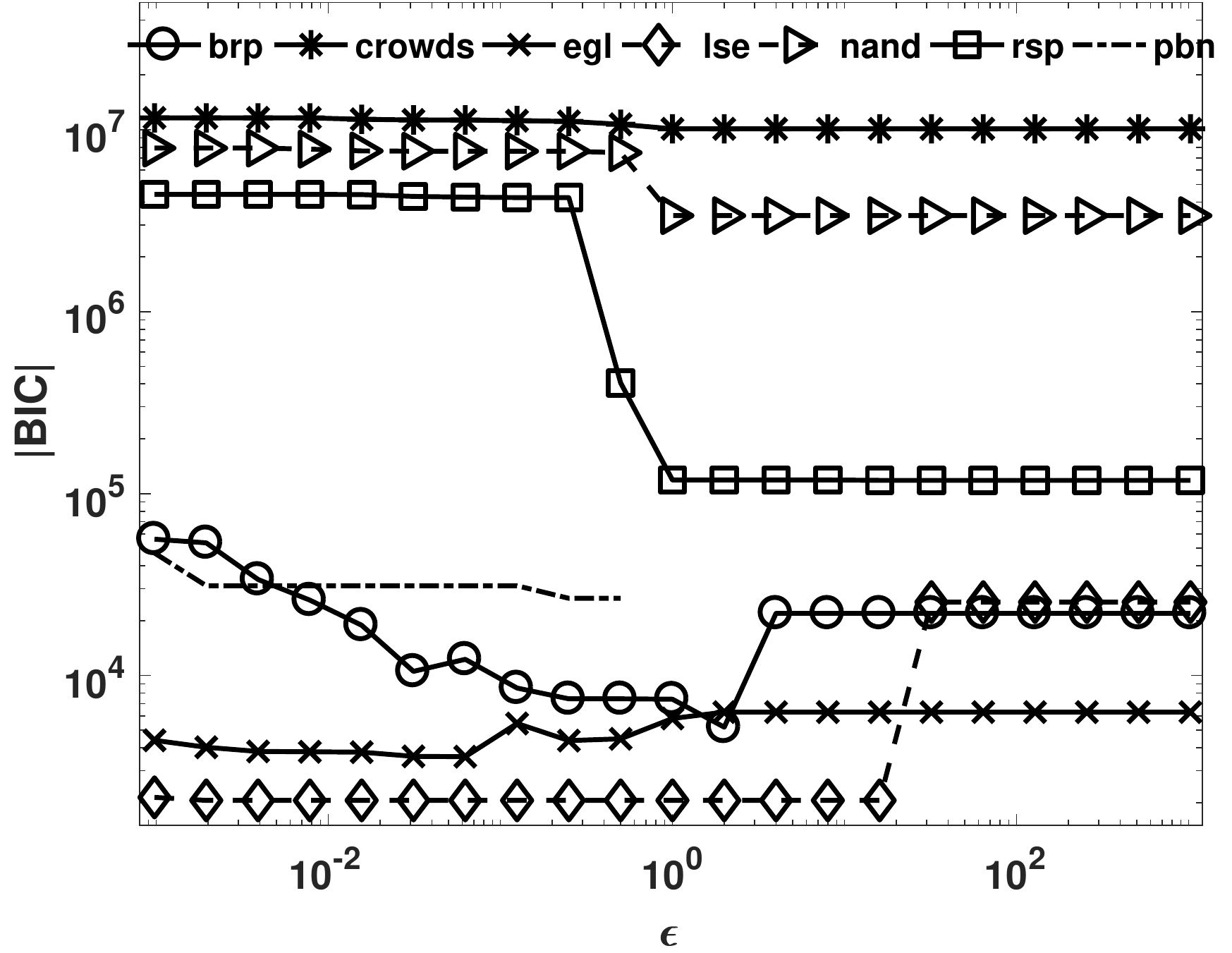}
 \caption{How the absolute values of BIC score change over $\epsilon$.}
 \vspace{-8mm}
 \label{fig:bic}
 \end{wrapfigure}

\emph{We first show that assumptions required by existing learning algorithms may not hold,} which motivates our proposal of GA-based algorithms. Existing learning algorithms~\cite{mao2011learning,chen2012learning} require that the BIC score is a concave function of $\epsilon$ in order to select the best $\epsilon$ value which controls the degree of generalization. Figure~\ref{fig:bic} shows how the absolute value of BIC scores ($|\textit{BIC}|$) of representative models change with $\epsilon$. It can be observed that this assumption is not satisfied and $\epsilon$ is not controlling the degree of generalization nicely. For example, the $|\textit{BIC}|$ (e.g., for \emph{brp}, \emph{PBN} and \emph{egl}) fluctuate with $\epsilon$. Besides, we observe climbings of $|\textit{BIC}|$ for \emph{lse} when $\epsilon$ increases, but droppings for \emph{crowds}, \emph{nand} and \emph{rsp}. What's worse, in the case (e.g., \emph{PBN}) of learning from a single execution, if the range of $\epsilon$ is selected improperly, it is very likely that an empty model (a tree only with root $\langle\rangle$) is learned.

\begin{figure}[t]
    \centering
    \begin{subfigure}[b]{0.48\textwidth}
        \includegraphics[width=\textwidth]{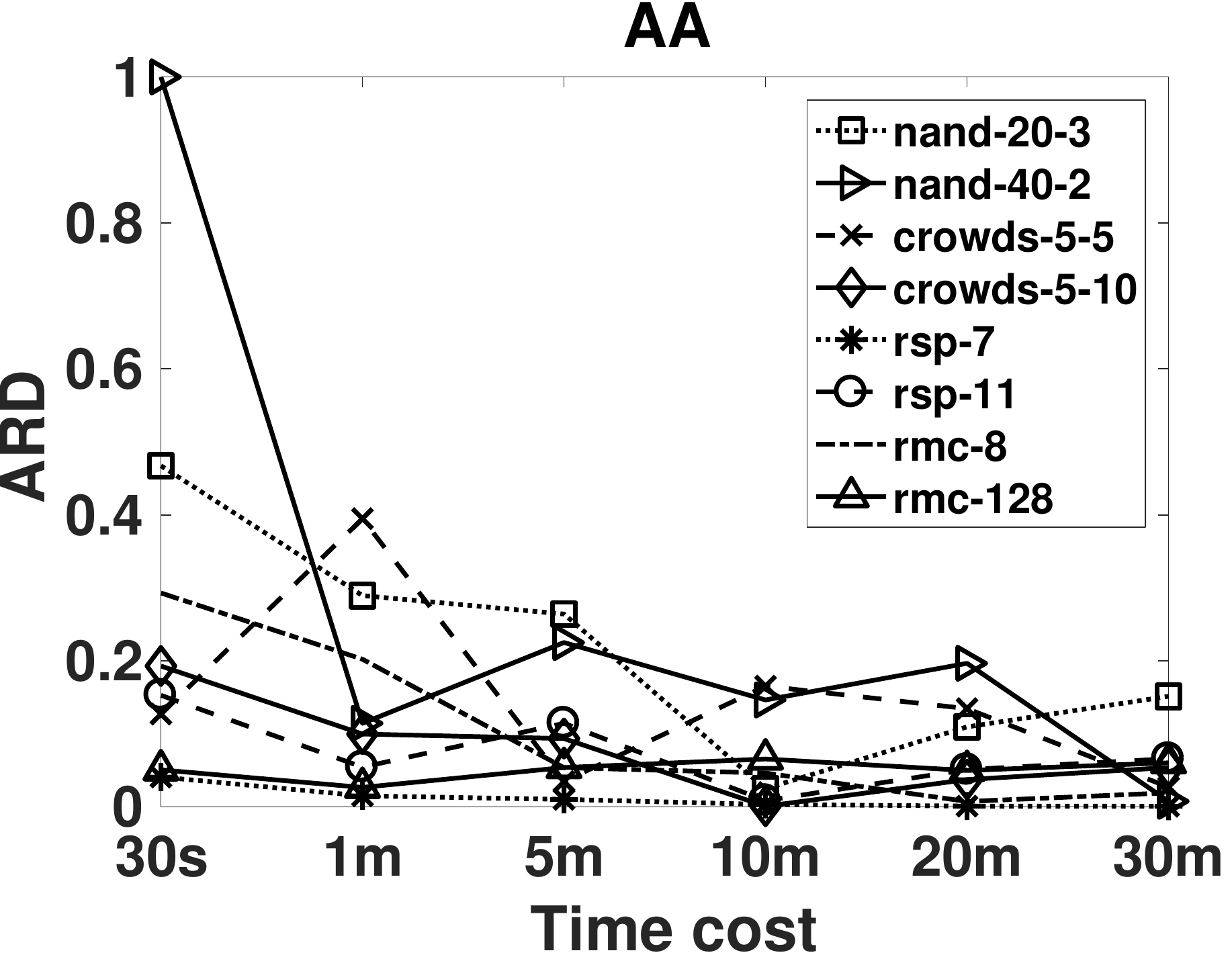}
        \label{fig:con:aa}
    \end{subfigure}
    ~ 
    \begin{subfigure}[b]{0.48\textwidth}
        \includegraphics[width=\textwidth]{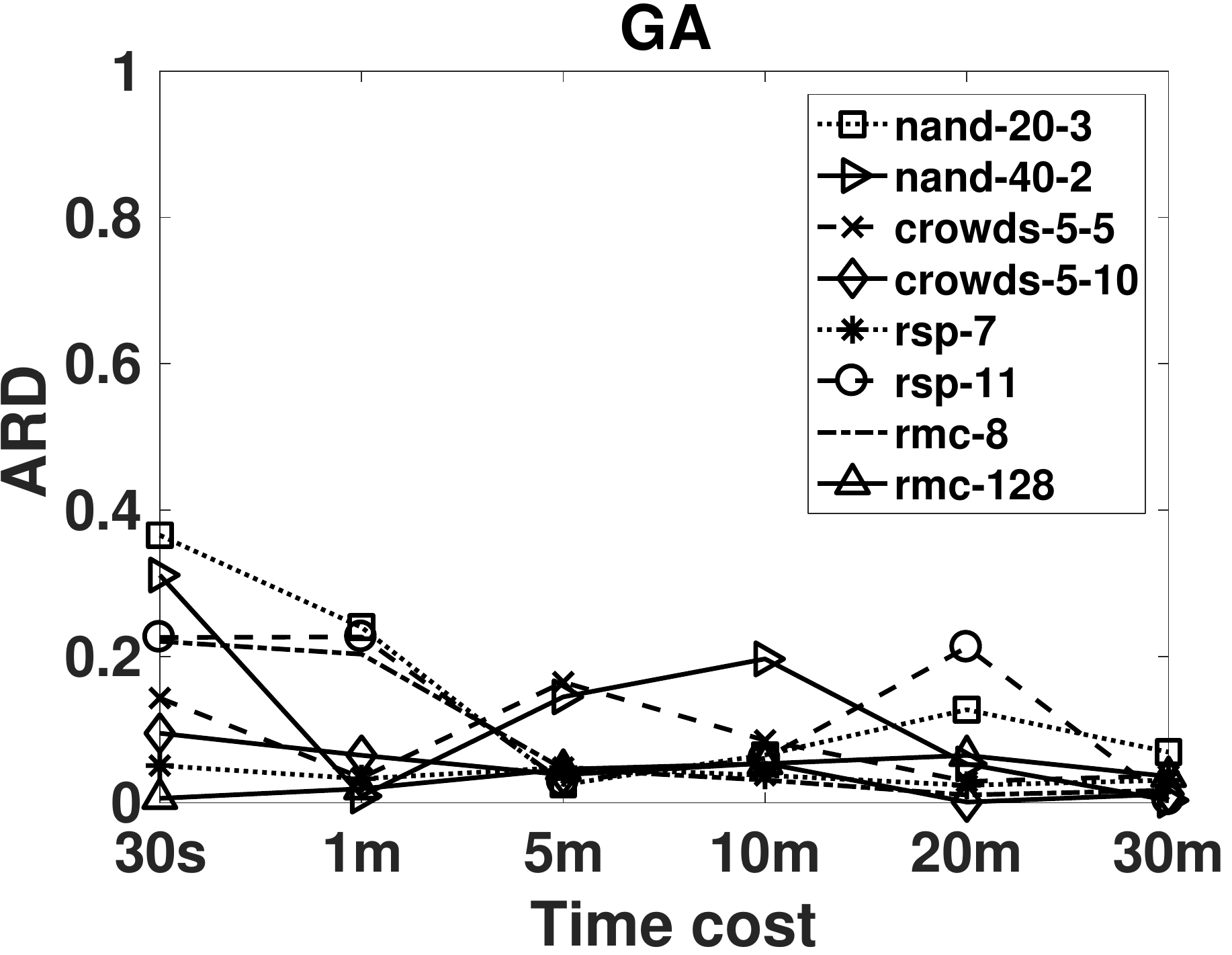}
        \label{fig:con:ga}
    \end{subfigure}
	\vspace{-4mm}
	\caption{Convergence of AA and GA over time. The numbers after the system of legends are one kind of system configuration.}
	\vspace{-4mm}
	\label{fig:conv}
\end{figure}

\emph{Second, how fast does learning converge?} In the rest of the section, we adopt absolute relative difference (ARD) as a measure of accuracy of different approaches. The ARD is defined as $|P_{est}-P_{act}|/P_{act}$ between the precise result $P_{act}$ and the estimated results $P_{est}$, which can be obtained by AA, GA as well as SMC. A smaller ARD implies a better estimation of the true probability. Figure~\ref{fig:conv} shows how the ARD of different systems change when we gradually increase the time cost from 30 seconds to 30 minutes by increasing the size of training data. We remark that some systems (\emph{brp}, \emph{egl}, \emph{lse}) are not applicable due to different reasons. We can observe that GA converges faster and better than AA. In general, both AA and GA converges to relatively accurate results when we are given sufficient time. But there are also cases of fluctuation of ARD, which is problematic, as in such cases, we would not know which result to trust (given the different verification results obtained with different number of sampled executions), and it is hard to decide whether we have gathered enough system executions for reliable verification results.   

\begin{figure}[t]
    \centering
    \begin{subfigure}[b]{0.32\textwidth}
        \includegraphics[width=\textwidth]{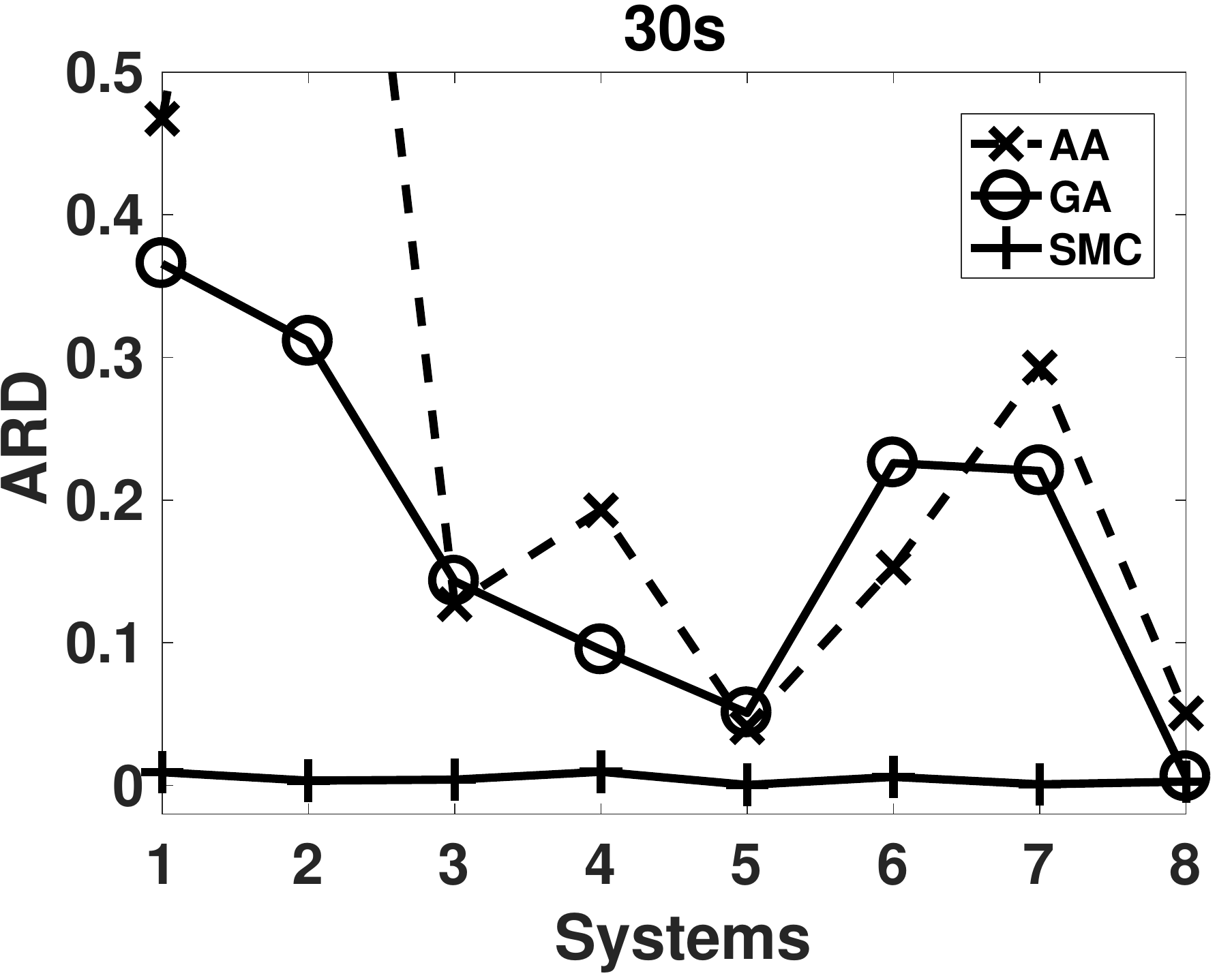}
        \label{fig:30s}
    \end{subfigure}
    ~ 
    \begin{subfigure}[b]{0.32\textwidth}
        \includegraphics[width=\textwidth]{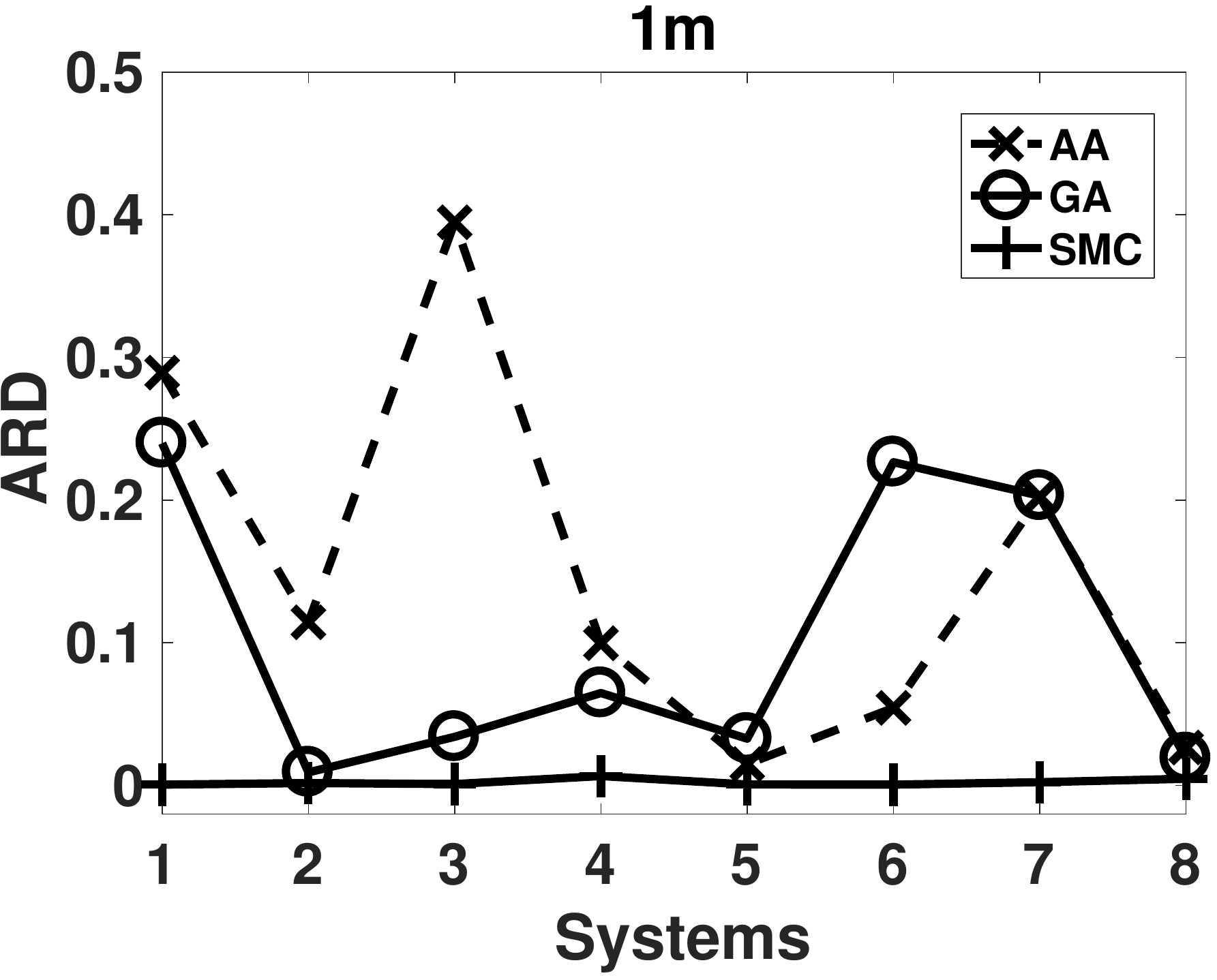}
        \label{fig:1m}
    \end{subfigure}
    ~ 
    \begin{subfigure}[b]{0.32\textwidth}
        \includegraphics[width=\textwidth]{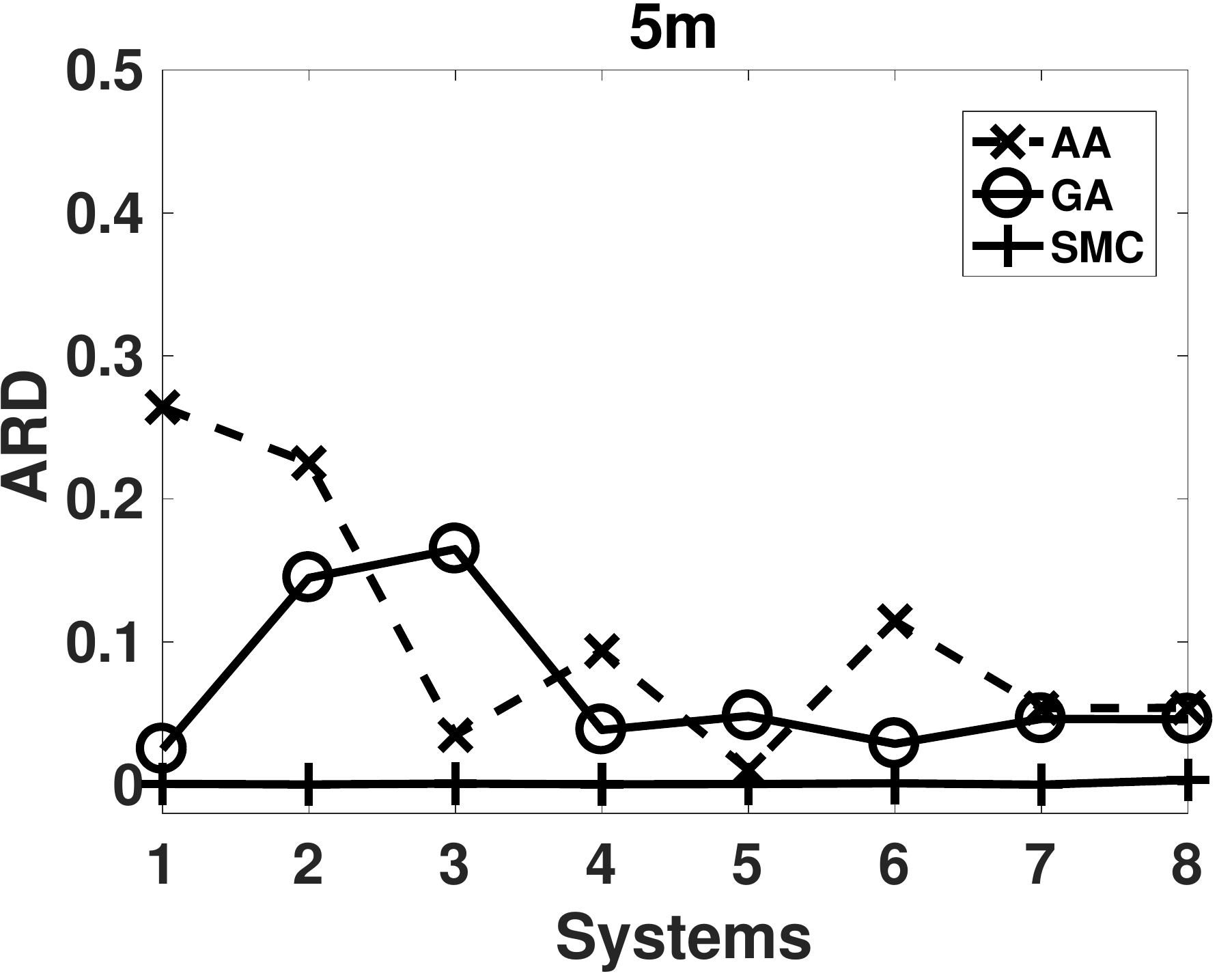}
        \label{fig:5m}
    \end{subfigure}
    \\
    \begin{subfigure}[b]{0.32\textwidth}
        \includegraphics[width=\textwidth]{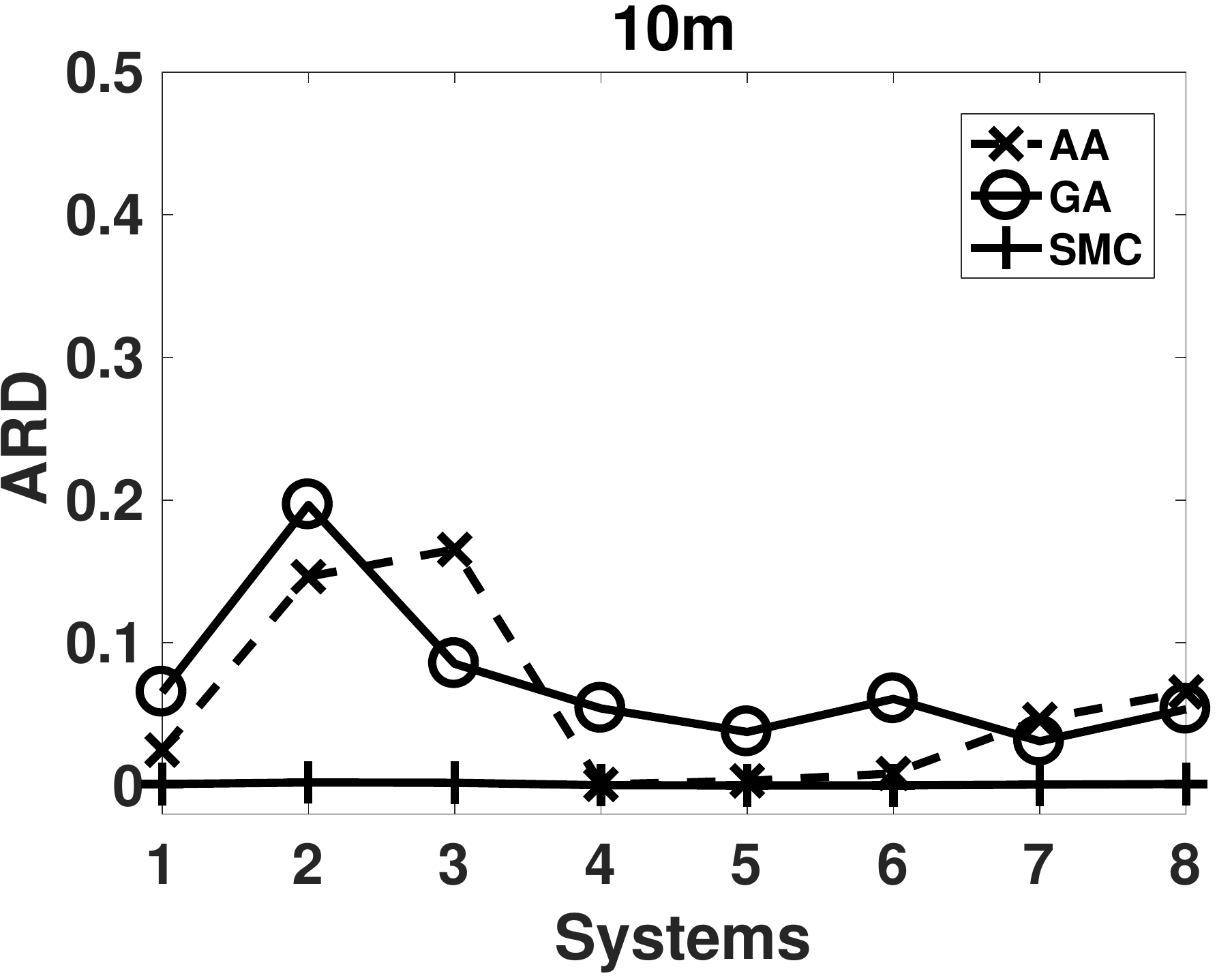}
        \label{fig:10m}
    \end{subfigure}
    ~ 
    \begin{subfigure}[b]{0.32\textwidth}
        \includegraphics[width=\textwidth]{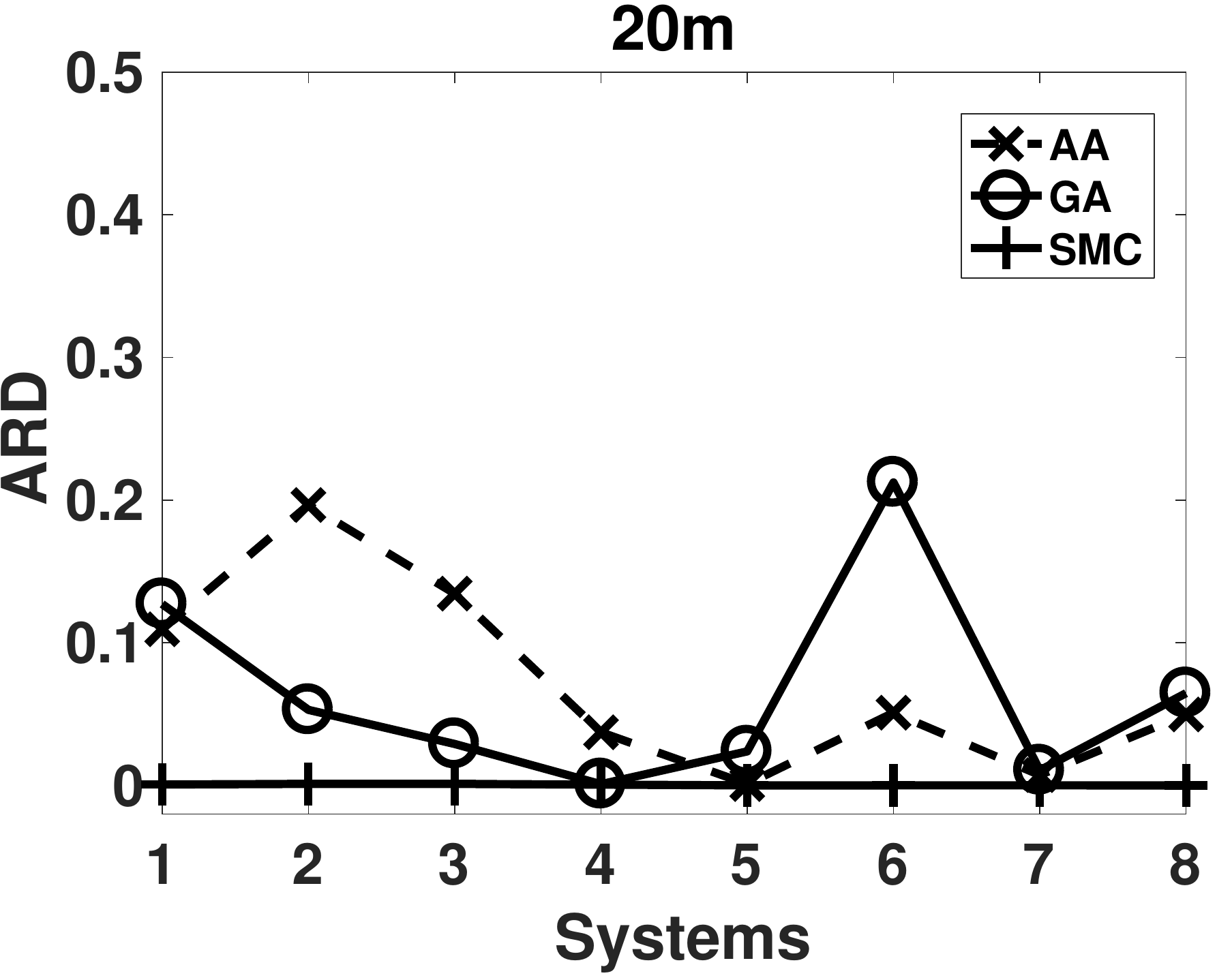}
        \label{fig:20m}
    \end{subfigure}
    ~ 
    \begin{subfigure}[b]{0.32\textwidth}
        \includegraphics[width=\textwidth]{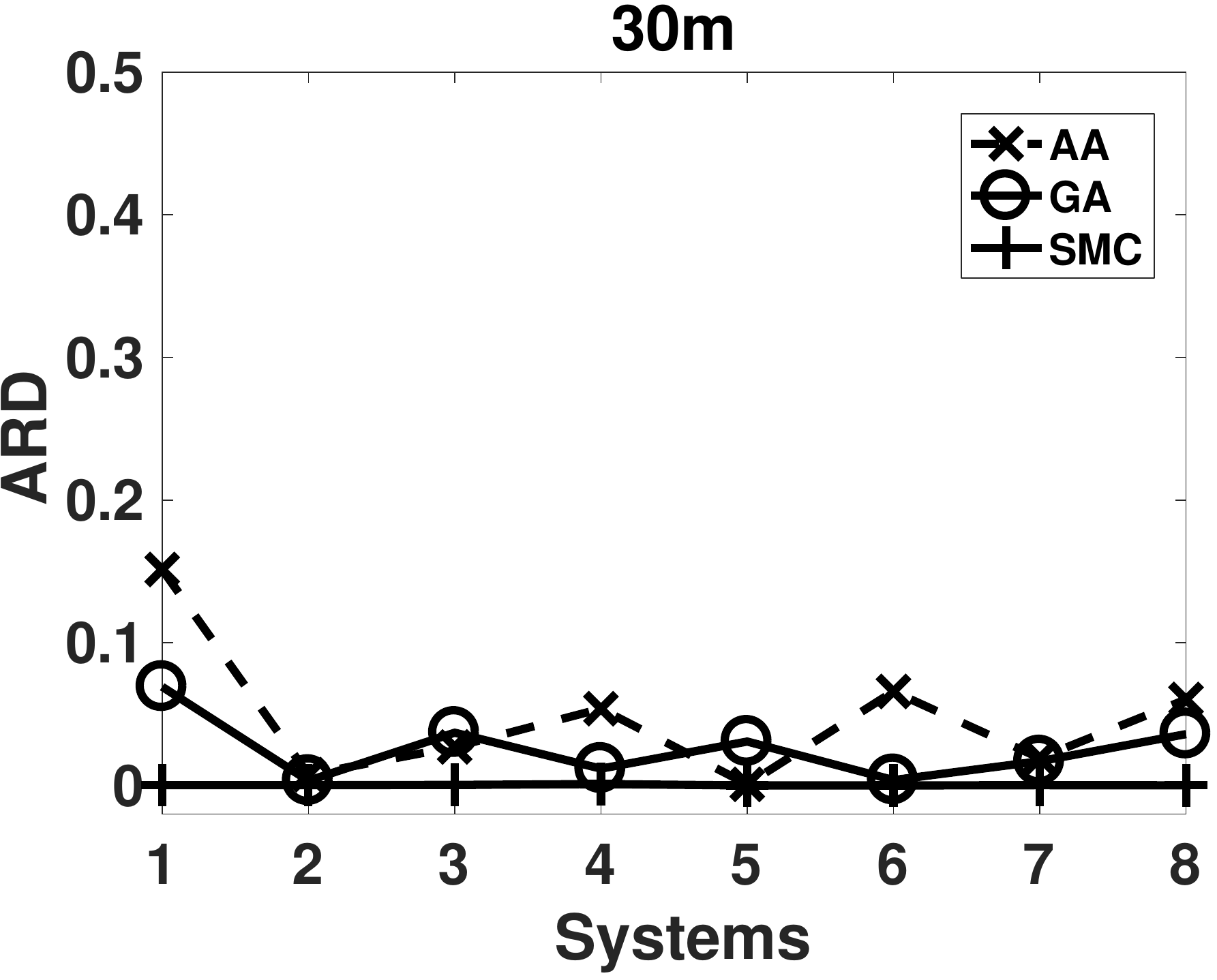}
        \label{fig:30m}
    \end{subfigure}
    \caption{The comparison of accuracy of AA, GA, and SMC given same amount of time, which varies from 30 seconds to 30 minutes. The horizontal-axis point represents a benchmark system with certain configuration in Fig.~\ref{fig:conv}.}\label{fig:acc}
\end{figure}

\emph{Third, how accurate can learning achieve?} We compare the accuracy of AA, GA, and SMC for benchmark systems given the same amount of time in Fig.~\ref{fig:acc}. We remark that due to the discrimination of system complexity (state space, variable number/type, etc.), different systems can converge in different speed. For SMC, we adopt the statistical model checking engine of PRISM and select the confidence interval method. We fix confidence to 0.001 and adjust the number of samples to adjust time cost. We have the following observations based on Fig.~\ref{fig:acc}. Firstly, for most systems, GA results in more accurate results than AA given same amount of time. This is especially true if sufficient time (20m or 30m) are given. \emph{However, it should be noticed that SMC produces significantly more accurate results}. Secondly, we observe that model learning works well if the actual model contains a small number of states. Cases like random models with 8 states (rmc-8) are good examples. For systems with more states, the verification results could deviate significantly (like nand-20-3, rsp-11).

Among our test subjects, \emph{PBN} and \emph{SWaT} are representative systems for which manual modelling is extremely challenging. Furthermore, SMC is not applicable as it is infeasible to sample the executions many times for these systems. We evaluate whether we can learn precise models in such a scenario. Note that since we do not have the actual model, we must define the preciseness of the learned model without referring to the actual model. For \emph{PBN}, following~\cite{SDZ02}, we use mean squared error (MSE) to measure how precise the learned models are. MSE is computed as follows: $MSE=\frac{1}{n}\sum_{i=1}^{n}(\hat{Y_i}-Y_i)^{2}$ where $n$ is the number of states in \emph{PBN} and $Y_i$ is the steady-state probabilities of the original model and $\hat{Y_i}$ is the corresponding steady-state probabilities of the learned model. We remark that the smaller its value is, the more precise the learned model is. Table~\ref{tab:ssl} shows the MSE of the learned models with for \emph{PBN} with 5, 8, and 10 nodes respectively. Note that AA and GA learn the same models and thus have the same MSE, while GA always consumes less time. We can observe the MSEs are very small, which means the learned models of \emph{PBN} are reasonably precise.

For the \emph{SWaT} system, we evaluate the accuracy of the learned models by comparing the predicted observations against a set of test data collected from the actual system. In particular, we apply steady-state learning proposed in~\cite{chen2012learning} (hereafter SL) and GA to learn from executions of different length and observe the trends over time. We select $3$ critical sensors in the system (out of 50), named $ait502$, $ait504$ and $pit501$, and learn models on how the sensor readings evolve over time. During the experiments, we find it very difficult to identify an appropriate $\epsilon$ for SL in order to learn a non-empty useable model. Our GA-based approach however does not have such problem. Eventually we managed to identify an optimal $\epsilon$ value and both SL and GA learn the same models given the same training data. A closer look at the learned models reveals that they are all first-order Markov chains. This makes sense in the way that sensor readings in the real \emph{SWaT} system vary slowly and smoothly. Applying the learned models to predict the probability of the test data (from another day with length 7000), we observe a very good prediction accuracy. We use the average prediction accuracy for each observation $\bar{P}_{obs}=P_{td}^{1/|td|}$, where $td$ is the test data and $|td|$ is its length, to evaluate how good the models are. In our experiment, the average accuracy of prediction for $ait502$ and $pit501$ is over 0.97, and the number is 0.99 for $ait504$, which are reasonably precise.

\begin{table*}[t]
\centering
\vspace{-2mm}
\caption{Results of \emph{PBN} steady-state learning.}
\label{tab:ssl}
\begin{adjustbox}{max width=\textwidth}
\begin{tabular}{cccccccccccc}
\toprule
\multirow{2}{*}{\# nodes}  &\multirow{2}{*}{\# states} & \multirow{2}{*}{\specialcell{trajectory \\ size ($\times 10^{3}$)}}  & \multicolumn{2}{c}{time cost(s)} &\multirow{2}{*}{\specialcell{MSE \\($\times 10^{-7}$) }}&\multirow{2}{*}{\# nodes}  &\multirow{2}{*}{\# states} & \multirow{2}{*}{\specialcell{trajectory \\size ($\times 10^{3}$)}}  & \multicolumn{2}{c}{time cost(s)} &\multirow{2}{*}{\specialcell{MSE \\($\times 10^{-7}$) }}\\ \cline{4-5}\cline{10-11}
&&& SL& GA &&&&&SL&GA&\\\midrule
\multirow{6}{*}{5} & \multirow{6}{*}{32} & 5 & 37.28 & 6.37 & 36.53 &\multirow{6}{*}{8}&\multirow{6}{*}{256}&5&29.76&2.36&1.07 \\
&&  15 & 161.57 & 53.49 & 15.21   &&& 15 & 105.87& 26.4 & 0.03   \\
&&  25 & 285.52 & 182.97 & 6.04   &&& 25 & 197.54& 73.92 & 0.37   \\
&&  35 & 426.26 & 348.5 & 7.75   &&& 35 & 310.87& 122.61 & 0.94  \\
&&  45 & 591.83 & 605.1 & 5.74   &&& 45 & 438.09& 429.81 & 0.78    \\
&   \multirow{10}{*}{}  &  50 &      673.55 & 767.7 & 4.28&&&50&509.59&285.66&0.34 \\ \midrule
\multirow{2}{*}{10}    &   \multirow{2}{*}{1024}  & 5 & 902.69& 266.74& 1.78 &\multirow{2}{*}{10} & \multirow{2}{*}{1024} & 15 & 5340.54&2132.68 & 0.61  \\
&& 10  & 2772.56&1010.16 & 1.01   &&& 20 & 8477.24& 3544.82& 0.47    \\ \bottomrule\hline
\end{tabular}
\end{adjustbox}
\vspace{-2mm}
\end{table*}

\emph{Last, there are some potential problems that may render learning ineffective.} One of them is the known problem of rare-events. For \emph{brp} system, the probability of satisfying the given properties are very small. As a result, a system execution satisfying the property is unlikely to be observed and learned from. Consequently, the verification results based on the learned models are $0$. It is known that SMC is also ineffective for these properties since it is also based on random sampling. Besides, learning doesn't work when the state space of underlying system is too large or even infinite. If there are too many variables to observe (or when float/double typed variables exist), which induces a very large state space, learning will become infeasible. For example, to verify the fairness property of \emph{egl} protocol, we need to observe dozens of integer variables. Our experiment suggests that AA and GA take unreasonable long time to learn a model, e.g., more than days. In order to apply learning in this scenario, we thus have to apply abstraction on the sampled system executions and learn from the abstract traces. Only by doing so, we are able to reduce the learning time significantly (in seconds) and successfully verified the \emph{egl} protocol by learning. However, how to identify the right level of abstraction is highly non-trivial in general and is to be investigated in the future. What's more, there are other complications which might make model learning ineffective. For the \emph{lse} protocol, the verification results based on the learned models may deviate from actual result for properties that show the probability of electing a leader in $L$ rounds, with a different value for $L$. While the actual result `jumps' as $L$ increases, the result based on the learned model is smooth and deviates from actual results significantly when $L$ is $3$, $4$ or $5$, while results based on SMC are consistent with the actual results.

\section{Conclusion and Related Work}\label{relatedwork}
In this work, we investigate the validity of model learning for the purpose of PMC. We propose a novel GA-based approach to overcome limitations of existing model learning algorithms and conducted an empirical study to systematically evaluate the effectiveness and efficiency of all these model learning approaches compared to statistical model checking over a variety of systems. We report their respective advantages and disadvantages, potential applications and future direction to improve.

This work is inspired by the work on comparing the effectiveness of PMC and SMC~\cite{younes2006numerical} and the line of work on adopting machine learning to learn a variety of system models (e.g., DTMC, stationary models and MDPs) for system model checking, in order to avoid manual model construction~\cite{mao2011learning,chen2012learning,mao2012learning}. Existing learning algorithms are often based on algorithms designed for learning (probabilistic) automata, as evidenced in~\cite{ron1996power,ron1995learnability,carrasco1994learning,de2010grammatical,carrasco1999learning,angluin1987learning}. Besides the work in~\cite{mao2011learning,chen2012learning,mao2012learning} which have been explained in detail, this work is also related to the work in~\cite{sen2004learning}, which learns continuous time Markov chains. In addition, in~\cite{brazdil2014verification}, learning algorithms are applied in order to verify Markov decision processes, without constructing explicit models. Our proposal on adopting genetic algorithms is related to work on applications of evolutionary algorithms for system analysis. In~\cite{he2010integrating}, evolutionary algorithm is integrated to abstraction refinement for model checking. This work is remotely related to work on SMC~\cite{younes2002probabilistic,sen2004statistical}, some recent work on extending SMC to unbounded properties~\cite{younes2011statistical,rohr2013simulative}. 
 Lastly, our work uses the PRSIM model checker as the verification engine~\cite{kwiatkowska2002prism} and the case studies are taken from various practical systems and protocols including~\cite{herman1990probabilistic,helmink1994proof,reiter1998crowds,norman2006analysis,norman2005evaluating,itai1990symmetry,assa}.

\bibliographystyle{splncs03}
\bibliography{paper1}
\end{document}